\documentclass[aps,prl,showpacs,floats,twocolumn,amsfonts,amssymb,unsortedaddress,floatfix]{revtex4}
\input epsf
\usepackage{bm}
\usepackage{amsmath}
\usepackage{graphicx}

\begin{document}

\addtolength{\topmargin}{10pt}

\def\Bbb{\mathbb}

\title{Thermomechanics  of DNA}

\author{Cristiano~Nisoli and A. R. Bishop}
\affiliation{\mbox{Theoretical Division and Center for NonLinear Studies, Los Alamos National Laboratory, Los Alamos NM 87545 USA}}

\date{\today}
\begin{abstract}

A theory for thermomechanical behavior of homogeneous DNA at thermal equilibrium predicts critical temperatures for denaturation under torque and stretch, phase diagrams for stable B--DNA, supercoiling, optimally stable torque, and the overstretching transition as force-induced DNA melting. Agreement with available single molecule manipulation experiments is excellent.  
  
\end{abstract}

\pacs{87.14.gk, 87.15.ad, 87.15.Zg,64.60.De}

\maketitle 

DNA is a highly refined nanomechanical object. The interplay between strong covalent bonds of the backbone and weak hydrogen interactions between bases \cite{Watson}, the thermal bath in which it is immersed, and the proximity of physiological conditions to the denaturation temperature, make DNA highly and non-linearly responsive to mechanical and thermal changes, and render any solely mechanical approach  inviable. This is critical for nanotechnology, where DNA is the basis for novel   materials~\cite{Seeman}, but understanding double helix thermomechanics would also illuminate biology, where enzymes involved in replication and repair are viewed as molecular motors. 

In the past twenty years, direct single molecule manipulation~\cite{Ritort, Bust} has revolutionized our understanding of key aspects of DNA, revealing new couplings and transitions between different structures, whose nature and forms, however, are still speculative. When a few micrometer long strand of DNA is stretched to a tension of the order of  pico-Newtons (pNs) to avoid  formation of plectonemes, sharp transitions are activated at positive and negative torques,  while an overstretching transition is observed for DNA under tension of $\simeq 60$ pN at zero torque~\cite{Bryant}. Tentative tension--torque phase diagrams for the stability of B--DNA, and various phenomenological theories, have been proposed to explain these effects. The robustness of the Peyrard-Bishop-Dauxois approach (PBD) \cite{PB,PBD, Dauxois} has been  corroborated by Cocco and Barbi~\cite{Cocco, Barbi}: they incorporated torque and successfully reproduced denaturation by unwinding. However, these recent studies do not include tension, do not explain denaturation at overwinding, and do not provide  phase diagrams in the tension-torque plane. Also, although much simpler than the molecular structure they describe, their complexity still calls for numerical treatments, and cannot offer analytic equations to more easily guide experiments.

We address these issues by modeling the steric dependence of the base bond and the effect of tension and torque in a way suitable for elimination of the angular degrees of freedom via integration of the partition function, and  obtain an effective energy for a PBD model which incorporates temperature and external loads. We compute the phase diagram for  B--DNA and the dependence of supercoiling on torque, tension and temperature at criticality. Finally, we propose   simple algebraic formul\ae~for the observables. 

In our  model  $i$ labels nucleotides separated by a distance $a$  along the DNA backbone (Fig. 1), $x_i$ is the length of the $i^{th}$ base's bond, $\omega_i=(\theta_{i+1}-\theta_{i-1})/2 -\Omega$ is the angular shift between nucleotides along the backbone, and $\theta_{i}$ is their angular coordinate. As in the Cocco--Barbi models the two strands of DNA are assumed symmetrical even when open. $\Omega$ denotes the natural pitch of the helix, and $\omega_i$ describes deviations from equilibrium.  The potential energy of the system is $E=a \sum_i E_i$ with 
\begin{equation}
E_i=\frac{k}{2} \frac{{\Delta x_i}^2}{a^2} +\frac{\nu}{2} \left(\omega_i+\Omega\right)^2 +\left[\chi(x_i)-1\right]V(\omega_i)
 \label{E}
\end{equation}
the sum of:  a stacking potential ($\Delta x_i=x_{i+1}-x_{i}$), in harmonic approximation for simplicity (see discussion later); an elastic term which restores the $\theta_{i+1}=\theta_{i} $ angular configuration for open strands; and a square well potential for the hydrogen bond between bases [$\chi(x)$ is a step function which is 0 for $0\le x \le x_c$ and 1 for $x_c <x$, where $x_c$ is a length associated with the hydrogen bond], whose depth $V$ depends on the angle $\omega_i$.  Because of a complex combination of hydrophobic, $\pi$--$\pi$, and dipolar interactions, the bonding of opposite bases is responsible for DNA's pitch. Therefore $V(\omega)$ is not symmetric but rather $V'(0)=\nu \Omega$, since DNA is in equilibrium at $\omega_i=0$.  Also, $\mu=-V''(0)$ must be positive: indeed, $\nu+\mu$ being  the torsional rigidity of the joined double helix and $\nu$ of the (much softer) open strands, we have $\mu \gg \nu>0$ (we shall see that $\mu/\nu\sim 10^2$). We can now expand $V$ as
\begin{equation}
V(\omega)\simeq V_0+\nu\Omega \omega-\frac{1}{2}\mu \omega^2.
\label{V}
\end{equation}
%

Our choice of a square well potential and separation of variables $x_i$, $\omega_i$ in (\ref{E}) keeps the model analytically solvable: no substantial changes in the thermodynamics arise from surrendering it in favor of, e.g., the Morse potential~\cite{Dauxois}. We show elsewhere how to introduce a smooth potential, desirable for dynamics and other studies~\cite{Nisoli2}. 

\begin{figure}[t!]   
\begin{center}
\includegraphics[width=1.7 in]{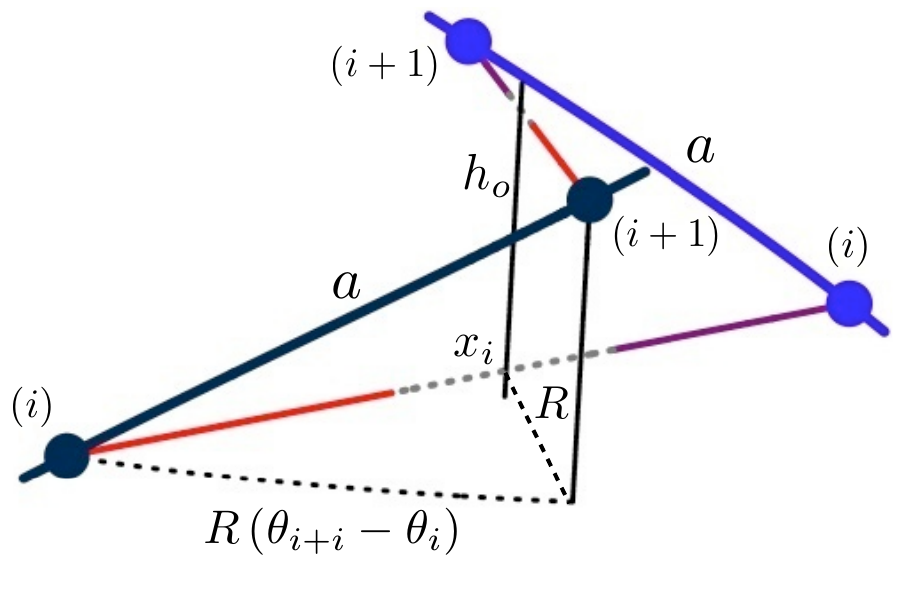}\hspace{3 mm}\includegraphics[width=.55 in]{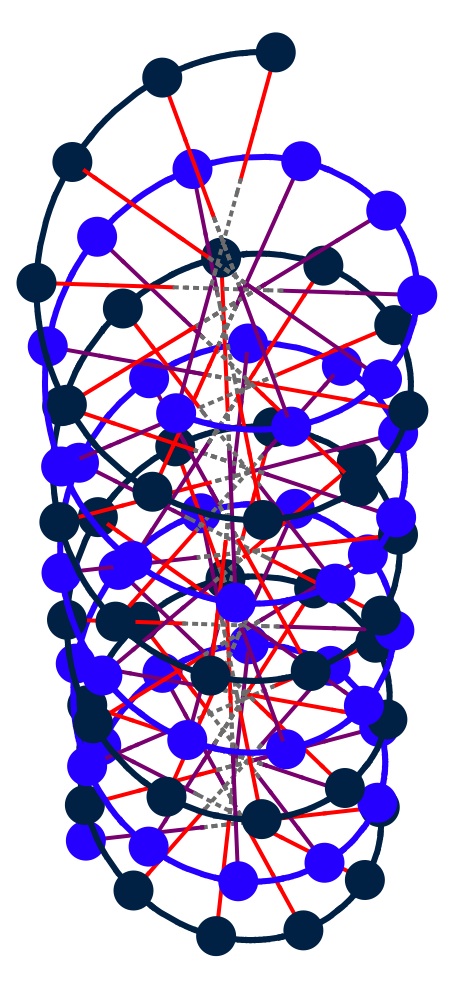}\hspace{5.5 mm}\includegraphics[width=.55 in]{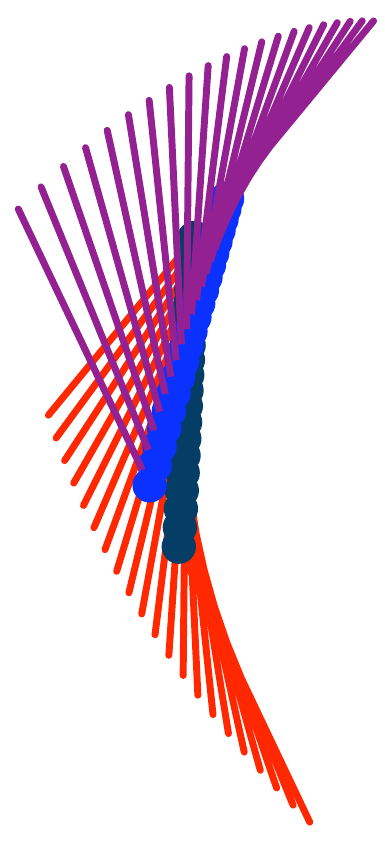}
\caption{Schematics of our DNA model showing degrees of freedom and geometrical parameters (Left), closed double helix (Middle), and open strands twisting with bases pointing outwards (Right). }
\end{center}
\label{draw}
\end{figure}

Now consider the external loads.  The torque $\Gamma$  is  incorporated in (\ref{E}) via a term -$\sum_i \tau \omega_i a$   (for dimensional convenience $\Gamma=\tau a$). Tension is more subtle: the total stretch $\sum_i h_i=\sum_i \chi(x_i) h_i^{o} + \sum_i\left[\chi(x_i)-1\right] h_i^{c}$ from both closed ($h_i^c$) and open ($h_i^o$) DNA sections can only arise from winding/unwinding, since the backbone is effectively inextensible. While elongation due to a change in pitch for the joined double helix is trivially $(h^c_i+h_0)^2=h_0^2-R^2 (\omega_i^2+2\omega_i \Omega)$ ($R$ is the radius of  the DNA helix, $h_0<a$ the vertical distance between nucleotides), for open sections assumptions are necessary.  We speculate that the two backbones twist with bases pointing outwards as in P-DNA (Fig.~1), and therefore the length of openings responds to winding: ${h^o_i}^2=a^2-r^2(\omega_i+\Omega)^2$, where $r<R$ is  an effective diameter  for the backbone. Below,  we will expand these expressions to second order around their stable configurations ($\omega_i=0$ for closed, $\omega_i=-\Omega$ for open sections).
All the quantities $E_i$, $k$, $\nu$, $\mu$, $V_0$, $\tau$, and $f$ have the dimension of a force.

Equilibrium thermodynamics is implemented by integration of $\exp{-\beta (E-\sum_i \tau \omega_i a -\sum_i f h_i)}$ over  $\{x_i\}$ and $\{\omega_i\}$. Since everything is quadratic in $\omega_i$, we can express the product of  the gaussian integrals in the angular variables $\omega_i$ and obtain the partition function
\begin{equation}
Z=e^{-\beta L\Delta} \int\prod_i dx_i e^{-\beta a\left[ \frac{k}{2} \frac{{\Delta x_i}^2}{a^2}+\tilde V(x_i)\right]}
\label{Z}
\end{equation}
for an equivalent PBD model, whose effective potential
\begin{equation}
\tilde V(x_i)=\left[\chi(x_i)-1\right]\left[\tilde V_0+ \tilde \Omega \tau -\frac{1}{2} \frac{\tilde \mu
}{(\tilde \nu+\tilde \mu)\tilde \nu} \tau^2 \right]
\label{Ecn}  
\end{equation}
incorporates explicitly the effect of the external torque $\Gamma= a \tau$, and  also of  $f$, through the tension-increased torsional rigidities $\tilde \mu=\mu+m f$, $\tilde \nu=\nu+n f$, and the pitch under tension $\tilde \Omega=\Omega-\frac{o}{\tilde\nu+\tilde \mu}f$. Also, the depth of the effective potential in the absence of external torque, 
$
\tilde V_0=V_0-\frac{T}{2a}\ln\frac{\tilde \nu+\tilde \mu}{\tilde \nu}-\frac{1}{2} \nu~\Omega^2-v f +\frac{\tilde \nu+\tilde \mu}{2}\left(\Omega-\tilde\Omega\right)^2
$,  
is weakened by tension $f$ and entropically by temperature $T$.  
The term $\Delta=-\frac{T}{2a}\ln\frac{2\pi T}{a \tilde \nu}-\frac{1}{2} \tilde \nu~\Omega^2-v f - \Omega \tau -\frac{1}{2\tilde \nu} \tau^2$ from (\ref{Z}), while irrelevant for the phase diagram, must be kept when computing  supercoiling. $L=N a$ is the length of the DNA. 
There are then purely geometrical  parameters: $m=\frac{R^2}{a h_0}+\frac{R^4}{a h_0^3}\omega_0^2-n$, $n=\frac{r^2}{a^2}$, $o=\frac{R^2}{a h_0}\Omega$, $v=1-\frac{h_0}{a}$, where $R~(\simeq10$ \AA)  is the radius of the DNA molecule,  $h_0~(\simeq3.4$ \AA) the elevation between consecutive nucleotides, $a~(\simeq 7$ \AA) their distance along the backbone, and $\Omega=2\pi/10$ the pitch of DNA: these are established geometrical values for B--DNA, but our formalism works, {\it mutatis mutandis}, for A-- and Z-- forms. 

\begin{figure}[t!!!!]   
\includegraphics[width=2.6 in]{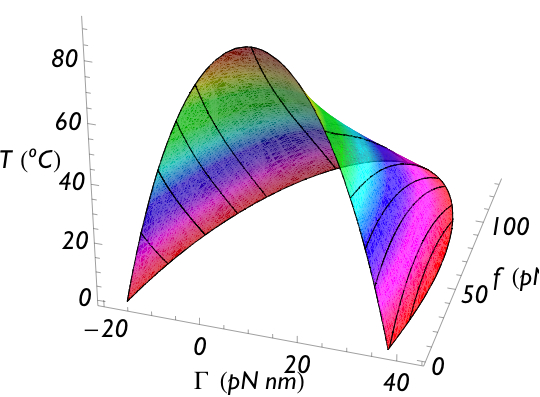}
\caption{Predicted critical surface for denaturation of B--DNA in ($T$), torque ($\Gamma$) and tension ($f$) obtained from (\ref{r}). Solid lines on the surface correspond to critical lines at fixed torque $\Gamma$ plotted in Fig.~\ref{Tf}.}
\label{Surface}
\end{figure}

The expression for $Z$ in (\ref{Z}) is exact within our model, and the resulting PB problem (or PBD if anharmonicity is included) is amenable to numerical transfer matrix treatment. Here we will  proceed analytically. In the continuum limit and neglecting an irrelevant equipartition factor, $Z$ in (\ref{Z}) can be written as
\begin{equation}
Z \propto  e^{-\beta L\Delta} ~\mathrm{Tr}~e^{-L \hat H},
\end{equation}
 proportional to the trace of the operator
\begin{equation}
\hat H=-\frac{1}{2k \beta}\partial_x^2+\beta \tilde V(x_i).
\label{H}
\end{equation}
Torque, tension and temperature enter the potential (\ref{Ecn}),  and the critical surface corresponds \cite{PB} to the disappearance of the bound state $\epsilon_B$ for $\hat H$, or 
\begin{eqnarray}
\tilde \Omega \tau -\frac{1}{2} \frac{\tilde \mu 
}{(\tilde \nu+\tilde \mu)\tilde \nu} \tau^2-v f +\frac{o^2 f^2}{\tilde\nu+\tilde \mu}+ \nonumber \\ 
+\frac{T_D^2-T^2}{\epsilon x_c}+\frac{1}{2a}\left(T_D l -T \tilde l\right)=0,
\label{r}
\end{eqnarray}
where $\epsilon=8kx_c/\pi^2$ has the dimension of an energy, $l=\ln[(\nu+\mu)/\nu]$ and $\tilde l=\ln[(\tilde\nu+\tilde\mu)/\tilde\nu]$ are pure numbers, and  $T_D^2=\epsilon x_c[V_0-\frac{T_D}{2a}\ln\frac{\nu+ \mu}{\nu}-\frac{1}{2} \nu~\Omega^2]$ is the denaturation temperature in the absence of torque or tension. 
\begin{figure}[t!]   
\includegraphics[width=3 in]{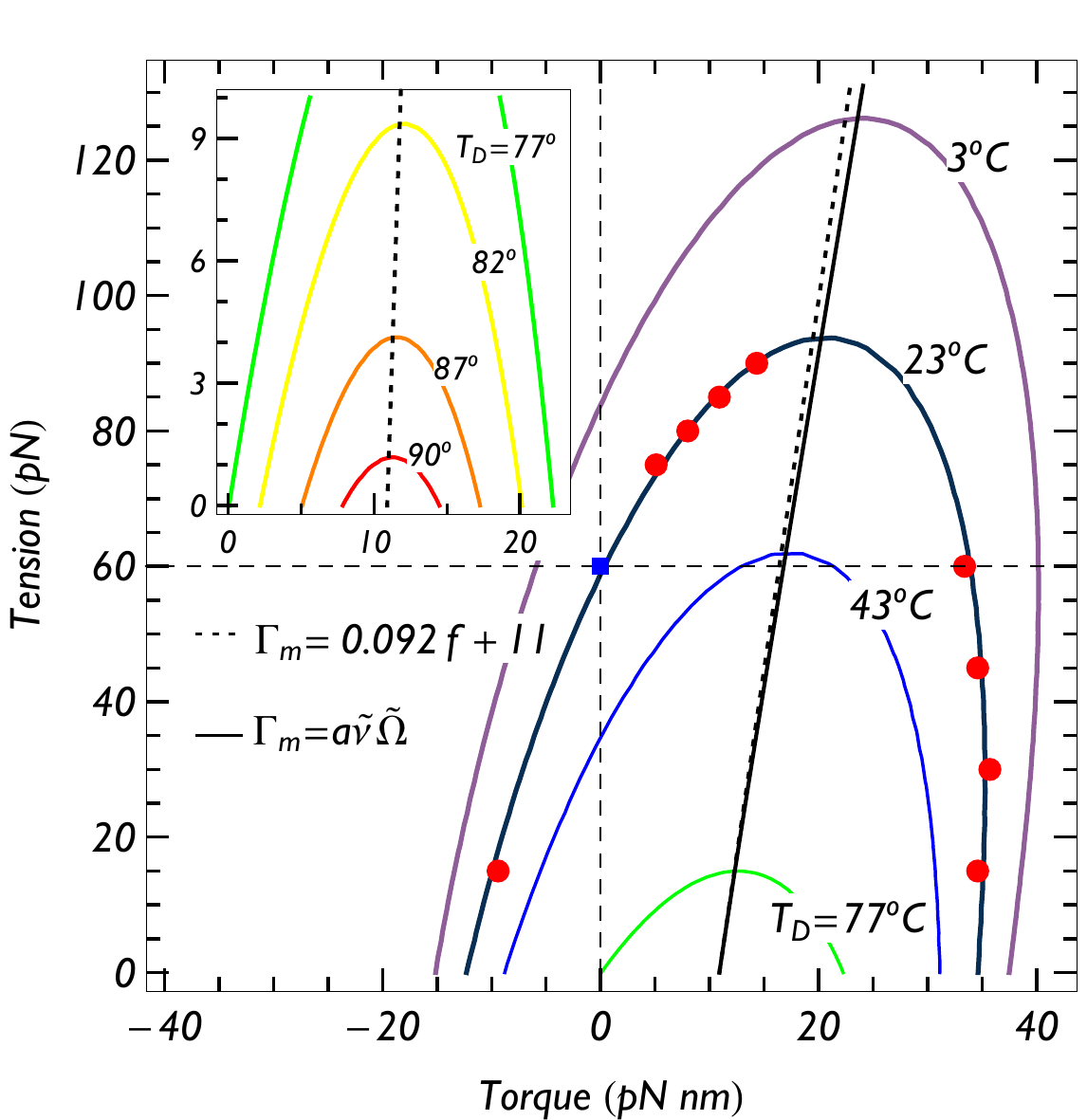}
\caption{Predicted critical lines for denaturation at different temperatures. The region enclosed by each line corresponds to stable B--DNA at that temperature. Points are experimental data from single molecule manipulation~\cite{Bryant}, square corresponds to the well known overstretching transition ($\Gamma=0$, $f=60$ pN). $T_D=350$ K $=77~^o$C, the denaturation temperature, is critical at zero external load: but even at temperatures of $T_D$ or higher (inset), B--DNA can be stable in an interval of applied positive torque. Solid (dotted) straight line indicates $\Gamma_m=a \tilde \nu \tilde \Omega$ (its approximation  $\Gamma_m= 11~+0.092 f $), the middle point between critical torques, which also corresponds to the highest critical temperature at given tension.}
\label{ft}
\end{figure}

Figure~\ref{Surface} shows our predicted critical surface of stable B--DNA in the space of $\Gamma$, $f$ and $T$.  There are only a few parameters to fit: an accepted value for the denaturation temperature is $T_D=350$ K~\cite{Saenger};  we show below from data on torque-winding experiments that $\mu=10^3$ pN. Choosing the remaining three parameters as  $\epsilon x_c=45$ pN\AA, $\nu=24$ pN and $n=0.3$, provides a remarkably good fit for 10 experimental data points (20 numbers)~\cite{Bryant} in the $f$ {\it vs.} $\Gamma$ phase diagram of Fig.~\ref{ft}.  The skewness of the critical lines can be quantified: from (\ref{r}) one finds for $\Gamma_m=(\Gamma_c^+ + \Gamma_c^-)/2$, the middle point between critical torques at  given tension,   
\begin{eqnarray}
\Gamma_m&=& a \frac{\tilde\mu+\tilde\nu }{\tilde\mu}\tilde \nu ~\tilde \Omega\simeq a\nu ~\Omega +f\left(n\Omega-\frac{o\nu}{\mu
}\right),
\label{Gm}
\end{eqnarray}
 independent of temperature. As (\ref{Gm}) reveals, skewness results from backbones twisting under torque: it would be erroneously negative for $r=0$. With the above parameters,
\begin{eqnarray}
{\Gamma_m}/\left(\mathrm{pN~nm}\right)&=11+92 \times 10^{-3}~{f}/\left(\mathrm{pN}\right).
\label{Gmh}
\end{eqnarray}
\begin{figure}[t!]   
\includegraphics[width=2.95 in]{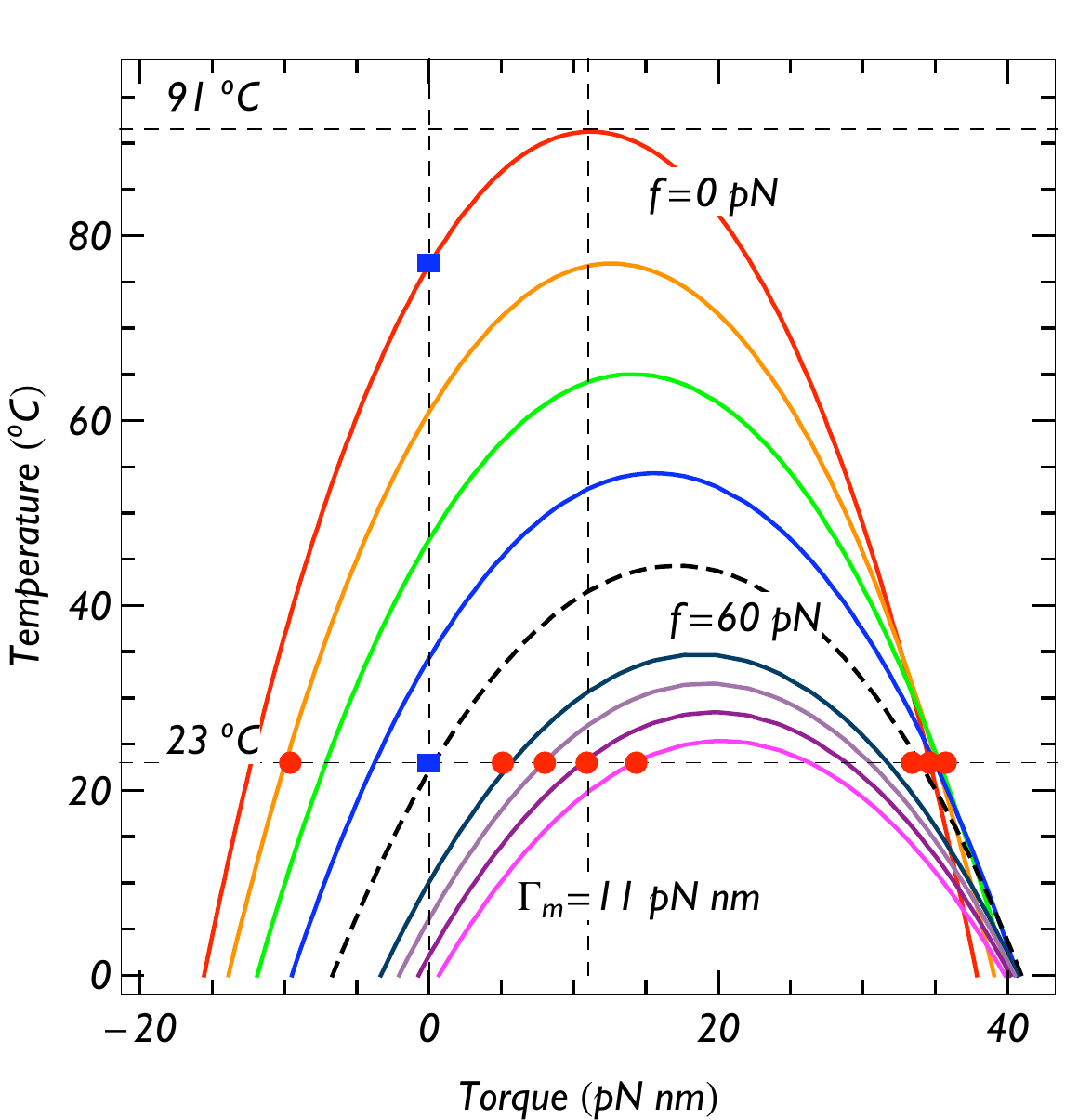}
\caption{Predicted critical temperature as function of torque for stretched DNA at different tensions (0,15, 30, 45, 60, 75, 80, 85 90 pN). Points are experimental data from single molecule manipulation~\cite{Bryant} performed at 23 $^o$C. Squares correspond to the overstretching transition ($\Gamma=0$, $f=60$ pN) at 23$^o$C and denaturation ($\Gamma=0$, $f=60$ pN) at $T_D=77~^o$C. The maximal critical temperature for a given stretch is achieved under external torque $\Gamma_m(f)$ given by~(\ref{Gm}), and is $91~^o$C at zero tension and $\Gamma_m=11$ pN nm.}
\label{Tt}
\end{figure}
\begin{figure}[t!!!!]   
\includegraphics[width=3 in]{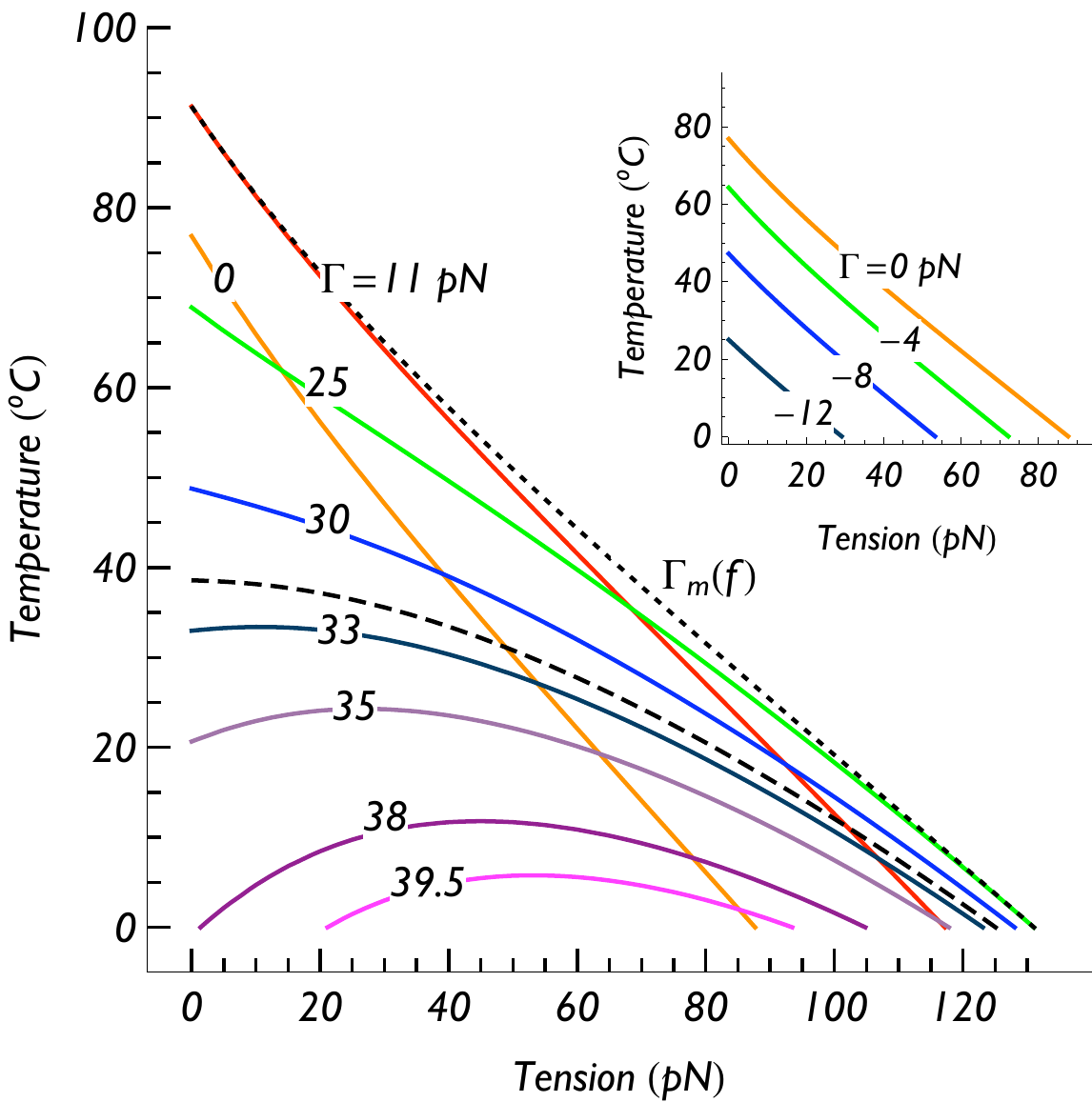}
\caption{Predicted critical temperature as a function of tension for DNA twisted under different torque. Numbers on curves denote the external torque in \mbox{pN nm}. The maximum critical temperature at any given tension (dotted curve) is reached at external torque $\Gamma_m(f)$ from (\ref{Gm}).  While for negative torque the curves show a monotonic and linear decrease (inset), for positive torque larger than 32 pN nm (dashed curve) curves become non-monotonic: for low temperatures and large applied torque, B--DNA is stabilized by tension.}
\label{Tf}
\end{figure}
%
Equations~(\ref{Gm}, \ref{Gmh}) along with experimental data for DNA that place melting at $f=15$ pN, for $\Gamma_c^+=34$ pN nm, $\Gamma_c^-=-10$ pN nm and at $f=60$ pN,  for $\Gamma_c^+=33$ pN nm~\cite{Bryant}, predict melting at zero torque and tension $f=60$ pN (blue square in Fig.~\ref{ft}) in good agreement with the observed  overstretching transition. Our analysis implies a force-induced melting~\cite{Williams} rather than a transition to a double helix with distortions~\cite{Ha}.
A positive torque stabilizes  DNA even at temperatures above denaturation (Fig~\ref{ft}, inset), a property exploited by thermophile bacteria living at high temperatures~\cite{Watson}. As expected, negative torque destabilizes DNA, a mechanism exploited in biology for DNA opening and replication. 

Under a given stretch, the critical temperature increases (decreases) under positive (negative) torque, as shown in Fig.~\ref{Tt}. From (\ref{r}) we see that  $T^c$ is maximized at $\Gamma=\Gamma_m(f)$, which therefore induces {\it the most stable configuration at any temperature}, for a certain tension $f$. The highest temperature B--DNA can withstand is achieved under zero tension and positive torque $\Gamma_m= a \nu \Omega=11$ pN nm, and corresponds to $T^c_m=91~^o$C.

DNA supercoils for packaging inside cells. This corresponds to small positive or negative torques; in that biological regime, the critical temperature decreases monotonically with an applied tension, in fact linearly, with slope independent of the applied torque or degree of supercoiling. Yet, for torques larger than about 32 pN nm, a regime accessible to single molecule manipulation experiments and potentially useful in nanotechnology, the maximal critical temperature corresponds to a non-zero tension, suggesting that at low temperatures and large torques  DNA can be stabilized by tension (Fig.~\ref{Tf}). 

The average change in pitch is given by $\langle \omega \rangle=L^{-1}\beta^{-1}\partial_{\tau} \ln Z=\tau/\tilde \nu -\Omega-T \partial_{\tau} \epsilon_B$  ($\epsilon_B$ is the bound state of the hamiltonian (\ref{H}), whereas the remainder comes from $\Delta$). At the critical point, we find
\begin{equation}
\langle \tilde \omega_c \rangle=\omega_D-\frac{\omega_D}{\Omega}\frac{\Gamma_c}{a\tilde \nu} +\left(1+\frac{\omega_D}{\Omega}\right)\frac{a^{-1}\Gamma_c-o f}{\tilde \mu+\tilde \nu}\simeq \frac{\Gamma_c}{a \tilde \mu},
\label{omega}
\end{equation}
where $\omega_D$ is the small negative unwinding (typically $\omega_D\simeq-10^{-4}$ rad)  at denaturation ($T=T_D$, $\Gamma=0$, $f=0$).  Equation (\ref{omega}) can be used to fit, from experimental data~\cite{Bryant}, $\mu=10^3$ pN. When neglecting  small non-linear corrections, (\ref{omega}) is also a good approximation for the torque vs.  pitch curve away from criticality.
While the derivation of (\ref{omega}) will be presented elsewhere~\cite{Nisoli2}, we note here that a well potential on an infinite half line in (\ref{H}) would generate no contribution from the bound state $\epsilon_B$ to  $\langle \tilde \omega \rangle$ at criticality, and thus give $\langle \omega_c \rangle=\frac{\Gamma_c}{a\tilde \nu}-\Omega$, the expected behavior for separated strands. The discontinuity at criticality in $\langle \omega \rangle$ seen  in experiments therefore comes from the finiteness of lateral entropy: when bases open in a  bubble, they cannot move infinitely far from each other: as tension and torque are applied and backbones twist, the radius of DNA provides a natural barrier for large $x$. This suggests that in biology, where, unlike in manipulations under tension, larger distances are available between bases in larger openings, the average change in pitch $\langle \omega \rangle$ might access larger values before complete strand separation occurs. 

Equation~(\ref{omega}) does not predict the tiny anomalous overwinding under tension~\cite{Bust2}. We believe this is a consequence of neglecting the bending modes, whose thermal fluctuation weakens the base's bonds.  As tension straightens DNA, it might reduce this effect by collapsing openings  and thus overwinding the structure. Conversely,  torque--induced overwinding reduces the fraction of open bases, thereby stiffening the bending modes and reducing the amplitude of their thermal fluctuations, resulting in elongation. We will show elsewhere~\cite{Nisoli2} how to incorporate these effects into our formalism to predict elongation under external loads, and also other phenomena relevant to biology, such as bending induced by a genetic defect, and effects of non-uniform gene sequence. Finally, a transfer integral numerical study~\cite{Nisoli2} of the anharmonic version of~(\ref{Z}) is needed to augment our predictions at temperatures closer to denaturation, where the effect of nonlinearity in the stacking potential is important for the transition order and precursors~\cite{AB}.

In summary, a range of predictions and phenomena for DNA thermomechanics (critical lines, phase diagrams, supercoiling under loads, optimally stable torque,  tension-induced stability at high torque)  that were inaccessible to previous models, or could only be covered partially and numerically, have been explained here within a unifying framework amenable to analytical treatment and further extensions and applications. We are grateful to B. Alexandrov and K. Rasmussen (LANL), S. Ares and J. Bois (MPIPKS Dresden),  C. Matek and A. Louis (Oxford) for stimulating conversations, to Amity Law (Harvard), Cynthia Reichhardt (LANL), and Paul Lammert (PSU) for helping with the manuscript.

This work was carried out under the auspices of the National Nuclear Security Administration of the U.S. Department of Energy at Los Alamos National Laboratory under Contract No. DEAC52-06NA25396.

\end{document}